\documentclass[10pt]{article}
\usepackage[twoside,pdftex,papersize={210mm,280mm}, total={16.8cm,23.6cm}, left=1.5cm, top=1.5cm, right=1.5cm, bottom=2.5cm, includehead=true]{geometry}
\usepackage{amssymb}
\usepackage[affil-it]{authblk}
\usepackage[fleqn]{amsmath}
\usepackage{lineno}
\usepackage{epsfig}
\usepackage{subcaption}
\usepackage{hyperref}
\usepackage{lineno}
\usepackage{upgreek}
\usepackage{rotating}
\usepackage{pdflscape}

\begin{document}

\title{Track reconstruction through the application of the Legendre Transform on ellipses}

\author{T.~Alexopoulos$^{1}$, Y.~Bristogiannis$^{1}$\footnote{hyploc@gmail.com}\ and S.~Leontsinis$^{2}$}

\affil{$^1$National Technical University of Athens}
\affil{$^2$University of Colorado, Boulder}

\maketitle

\begin{abstract}
We propose a pattern recognition method that identifies the common tangent lines of a set of ellipses.
The detection of the tangent lines is attained by applying the Legendre transform on a given set of ellipses.
As context, we consider a hypothetical detector made out of layers of chambers, each of which returns an ellipse as an output signal.
The common tangent of these ellipses represents the trajectory of a charged particle crossing the detector.
The proposed method is evaluated using ellipses constructed from Monte Carlo generated tracks.
\end{abstract}

\section{Introduction}
\label{sec:intro}

The Legendre transform is a mathematical tool with many applications in classical mechanics~\cite{paper_legendre_sense},~\cite{paper_legendre_graph_deriv}, statistical mechanics, thermodynamics~\cite{paper_legendre_thermodynamics_calc},~\cite{paper_legendre_thermodynamics_venn}, and computer vision problems~\cite{paper_legendre_fenchel_computer}.
It can be used to get all the possible tangents from a curve that is either convex or concave.
The basis of the track reconstruction algorithm we propose is the transformation of each ellipse to the Legendre space where an intersection of two curves represents a common tangent of the transformed functions.

As shown in previous work by T. Alexopoulos et al~\cite{paper_legendre_circles},~\cite{paper_legendre_circles_det}, the Legendre transform can be applied on circles to detect their common tangent and in turn, the trajectory of a charged particle passing through the Monitored Drift Chamber detector of ATLAS~\cite{atlas_cern} experiment at CERN~\cite{lhc_cern}.
In this study, we identify the common tangent lines of a given set of ellipses.
The Legendre transform is the core of the the proposed method.
Each ellipse is transformed into the Legendre space, where the intersections of the transformed curves represent the common tangent lines of the ellipses.

To evaluate our method, we implement a hypothetical detector made of multiple layers of chambers whose output signal is an ellipse.
Based on these ellipses, we detect the common tangent line which in turn, represents the track of a particle transversing the detector.
Monte Carlo generated tracks that pass through the detector are produced.
As a track passes through the chambers, elliptical hits are created.
These ellipses are created so that they are concentric with the chambers and tangent to the track.
We consider events during which only one track passes through the detector, as well as events with multiple tracks at once.
Our method is studied against non-ideal conditions such as the smearing of the ellipses' dimensions 
and noise induced ellipses.

A binary status of success is set based on the difference between the characteristics of the initial and reconstructed tracks. The characteristics of a line, or track, are it's $\mathrm{slope}$ and it's $\mathrm{intercept}$ with the $y$-axis. The efficiency of the algorithm is evaluated as the ratio of the number of successfully reconstructed tracks to the number of Monte Carlo generated tracks.

\section{The Legendre Transform}
\label{sec:lt_intro}

The Legendre transform $F\left(p\right)$ for a value $p={\mathrm{d}f}/{\mathrm{d}x}$ of a given convex function $f:\mathbb{R}\rightarrow\mathbb{R}$, where ${\mathrm{d}^2 f}/{\mathrm{d}x^2} > 0$, is defined as:
$$
F\left(p\right)=\underset{x}{\sup}\left[p  x - f\left(x\right)\right]
=-\underset{x}{\inf}\left[f\left(x\right)-p  x\right]
,$$
and that of a concave function, where ${\mathrm{d}^2 f}/{\mathrm{d}x^2} < 0$, as: 
$$
F\left(p\right)=\underset{x}{\sup}\left[f\left(x\right)-p  x\right]
=-\underset{x}{\inf}\left[p  x - f\left(x\right)\right]
.$$

\begin{figure}[h]\centering
         \includegraphics[scale=.50]{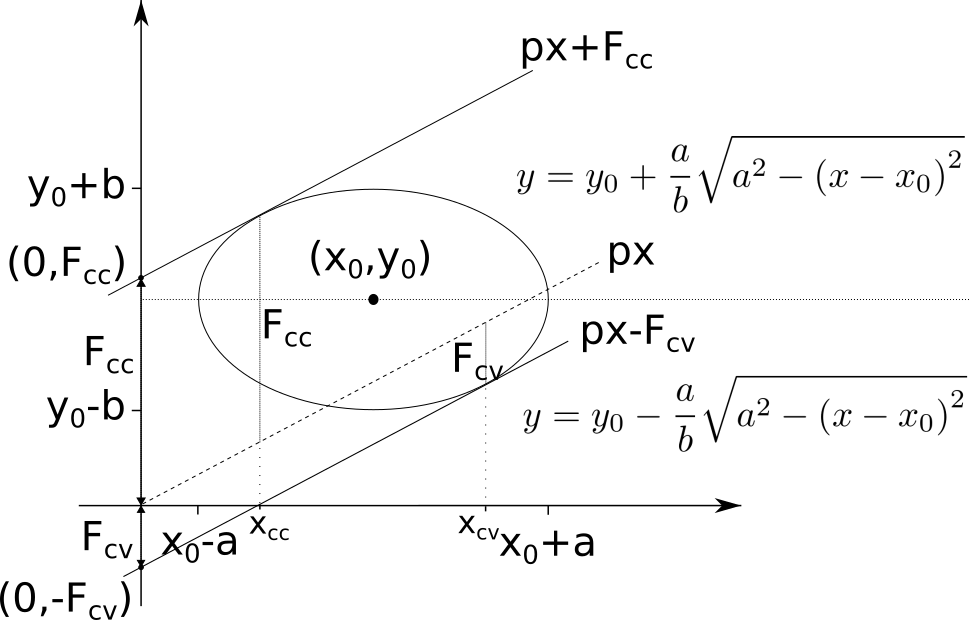}
        \caption{Representation of an ellipse by a concave (for $y > y_{0}$) and a convex (for $y < y_{0}$) function. The lines  $ y= px+ F_\mathrm{cc}$ and $ y= px+ F_\mathrm{cv}$ are examples of the Legendre transform for the concave and the convex part respectively. }
	\label{fig:ellipse_functions_lt_merged}
\end{figure}

The $\sup_x $ notation stands for the supremum of $p  x - f\left(x\right)$ with respect to $x$. 
In other words, we seek the maximum of the function $p  x - f\left(x\right)$ with respect to $x$, where $p$ is constant. 
Geometrically, this means that we have a function $f\left(x\right)$ and we are searching for $x$ so that a line with $\mathrm{slope} = p$ passes through $ \left( x, f\left(x\right) \right)$ and has a maximum intercept on the $y \mbox{-axis} $. 
That line will be tangent to $f\left(x\right)$ and nothing else since $f\left(x\right)$ is a convex function (figure~\ref{fig:ellipse_functions_lt_merged}). 
The $ \inf_x $ stands for infimum and therefore minimization or in other words, the minimum intercept with the $y \mbox{-axis} $.

If we solve $p={\mathrm{d}f}/{\mathrm{d}x}$ for $x$, the Legendre transform can be expressed as a function of $p$:
\begin{equation}
\label{eq:lt}
F\left(p\right) = \begin{cases}
f\left(x\left(p\right)\right) - p  x\left(p\right) & \text{for a concave function}\\
p  x\left(p\right) - f\left(x\left(p\right)\right) & \text{for a convex  function}
\end{cases}
,\end{equation}
where each point $\left(p,F\left(p\right)\right) $ in the Legendre space represents a line, tangent to the curve $f\left(x\right)$ where the slope is $p$ and the intersection of the line with the y-axis is $-F\left(p\right)$ or $F\left(p\right)$ if the line is tangent to a convex function or a concave function respectively. Therefore, the linear equation is defined as:
\begin{equation}
\label{eq:lt_line}
y = \mathrm{slope} \cdot x + \mathrm{intercept} =
\begin{cases}
px\left(p\right) + F\left(p\right) & \text{for a concave function}\\
px\left(p\right) - F\left(p\right) & \text{for a convex  function}
\end{cases}
.\end{equation}

It is now clear that an intersection $\left(p_\mathrm{t},F\left(p_\mathrm{t}\right)\right) $ of two or more curves in the Legendre space represents a common tangent of the functions transformed into the Legendre space. This is the basis of the tangent line finding method that we propose.

\subsection{Transformation of the ellipse into the Legendre space}
\label{sec:ellipse}

Our method is based on the transform of the ellipse to the Legendre space.
So, before the description of our method, we will apply the Legendre transform on the ellipse and manipulate it to reduce the computational cost.

The equation of an ellipse with center $\left(x_0,y_0\right)$, major axis $a$, and minor axis $b$ is:
\begin{equation}
\label{eq:ellipse}
\frac{\left( x - x_{0} \right)^{2}}{a^{2}} + \frac{\left( y - y_{0} \right)^{2}}{b^{2}} = 1
,\end{equation}
where $ a > b > 0 $.

Since the Legendre transform is only applicable to functions that are either convex or concave, we define the ellipse as a combination of convex and concave functions. Therefore, equation~\ref{eq:ellipse} solved for $y$ is:
$$
y = y_{0} \pm \frac{b}{a}\sqrt{a^{2} - \left(x-x_{0}\right)^{2}}
.$$

Now, we can define the functions $f_{1} (x)$ and $f_{2} (x)$ that refer to the concave and convex part of the ellipse respectively.
$$
f(x)=\begin{cases}
f_{1}\left(x\right)=y_{0}+\frac{b}{a}\sqrt{a^{2}-\left(x-x_{0}\right)^{2}} & \text{for the concave part}\\
f_{2}\left(x\right)=y_{0}-\frac{b}{a}\sqrt{a^{2}-\left(x-x_{0}\right)^{2}} & \text{for the convex  part}
\end{cases}
.$$

With the equation of the ellipse broken down to a concave and a convex function, the application of the Legendre transform is trivial (figure~\ref{fig:ellipse_functions_lt_merged}).

For the concave part of the ellipse, the Legendre transform for a concave function is applied:
$$
F_{1}\left(p\right)=\underset{x}{\sup}\left[f_{1}\left(x\right)-px\right]
,$$
where $p$ is the first derivative of the function $f_{1}\left( x \right)$ and so:
$$
p=\frac{\mathrm{d}f_{1}}{\mathrm{d}x}=\frac{b}{a}\frac{-\left(x-x_{0}\right)}{\sqrt{a^{2}-\left(x-x_{0}\right)^{2}}}\Rightarrow x=x_{0}\pm\frac{a^{2}|p|}{\sqrt{b^{2}+a^{2}p^{2}}}
,$$
but since $f_{1}\left( x \right)$ is a concave function, we get:
$$
\begin{cases}
x > x_{0} & \mbox{if } p < 0
\\
x < x_{0} & \mbox{if } p > 0
\end{cases}
\Rightarrow
x=x_{0}-\frac{a^{2}p}{\sqrt{b^{2}+a^{2}p^{2}}}
.$$

Therefore the Legendre transform of the concave part of the ellipse is:
$$
F_{1}\left(p\right) = y_{0}-px_{0}+\sqrt{b^{2}+a^{2}p^{2}}
.$$

\begin{figure}[h]\centering
         \includegraphics[scale=.55]{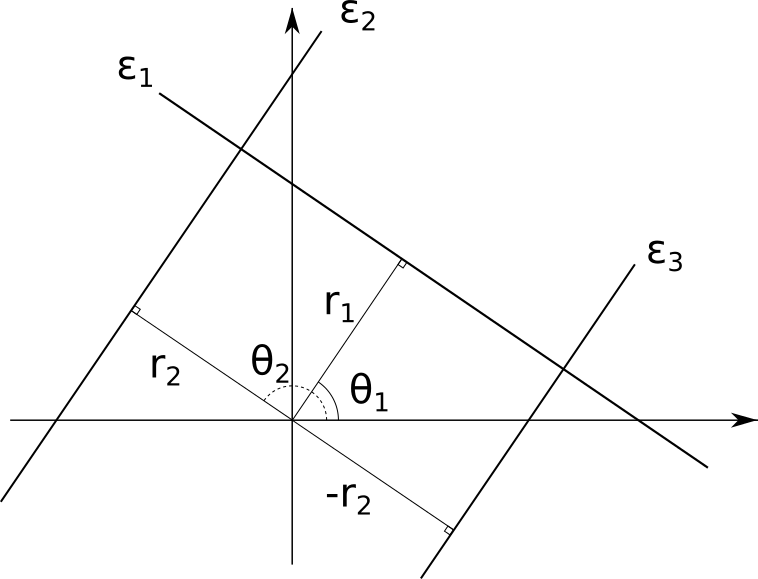}
        \caption{Lines and their canonical form. Line $\varepsilon_{1} $ has a canonical representation of a line segment that starts from the axis origin and is perpendicular to the line $\varepsilon_{1} $. The line segment of length $r_{1}$ and forms a $ \theta_{1}$ angle with the x-axis. Similarly, line $\varepsilon_{2} $ has a canonical representation of a line segment of length $r_{2}$ and slope $\tan \theta_{2}$. Line $\varepsilon_{2} $ is parallel to $\varepsilon_{2}$ and has the opposite intercept with $y$-axis as $\varepsilon_{2}$ has. Therefore $\varepsilon_{3} $ has a canonical representation of a line segment of length $-r_{2}$ and slope $\tan \theta_{2}$. The minus sign means that the intersection takes place below the $x$-axis.}
	\label{fig:canon_line}
\end{figure}

Though a tangent of the ellipse can be defined by a $\left( p , F_1 \left( p \right) \right) $ pair, due to the large span of values that $p$ can take, it is more suitable to express the linear equation by its canonical form (figure~\ref{fig:canon_line}), where $\theta \in \left( 0, \pi \right) $:
$$
r=x\cos\theta+y\sin\theta
,$$
so  that $p=-\cot\theta$ and $F\left(p\right)={r}/{\sin\theta}$.

Therefore, the Legendre transform becomes:
\begin{equation}
\label{eq:lte_concave}
\frac{r}{\sin\theta}=y_{0}+\frac{\cos\theta}{\sin\theta}x_{0}+\sqrt{b^{2}+a^{2}\left(\frac{\cos\theta}{\sin\theta}\right)^{2}}\Rightarrow
r=x_{0}\cos\theta+y_{0}\sin\theta+\sqrt{b^{2}\sin^{2}\theta+a^{2}\cos^{2}\theta}
.\end{equation}

Equation~\ref{eq:lte_concave} represents a sinogram in the $\left( \theta ,r \right)$ Legendre transformation space where each $\left( \theta ,r \left( \theta \right) \right)$ pair represents a line in the $\left( x ,y \right)$ space.

Following the same calculation steps, we conclude that the Legendre transform for the convex case is:
$$ F_{2}\left(p\right) = px_{0}-y_{0}+\sqrt{b^{2}+a^{2}p^{2}}
,$$ while each of the $\left( p , -F_2 \left( p \right) \right) $ pairs represent a tangent line in the $\left( x ,y \right)$ space. And in canonical form:
$$
r=x_{0}\cos\theta+y_{0}\sin\theta-\sqrt{b^{2}\sin^{2}\theta+a^{2}\cos^{2}\theta}
.$$

\begin{figure}[h]\centering
         \includegraphics[scale=.42]{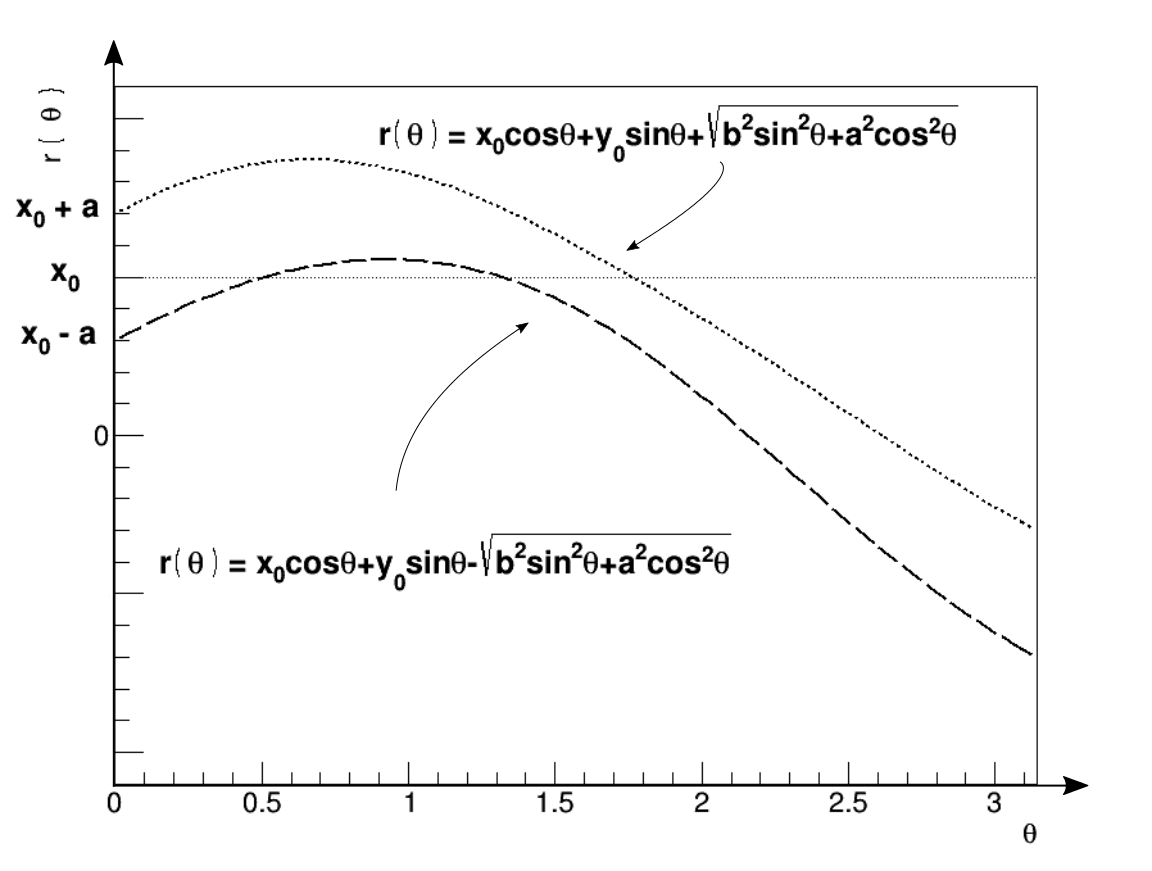}
        \caption{Representation of an ellipse in Legendre space.}
	\label{fig:lte_canon}
\end{figure}

Therefore, the Legendre transform of the ellipse is:
\begin{equation}
\label{eq:lte}
f\left(x\right)\leftrightarrow F\left( p \right) = \begin{cases}
y_{0}-px_{0}+\sqrt{b^{2}+a^{2}p^{2}} & \text{for the concave part}\\
px_{0}-y_{0}+\sqrt{b^{2}+a^{2}p^{2}} & \text{for the convex part}
\end{cases}
,\end{equation}
and in canonical form (figure~\ref{fig:lte_canon}):
\begin{equation}
\label{eq:lte_canon_form}
f\left(x\right)\leftrightarrow r\left( \theta \right) = \begin{cases}
x_{0}\cos\theta+y_{0}\sin\theta+\sqrt{b^{2}\sin^{2}\theta+a^{2}\cos^{2}\theta} & \text{for the concave part}\\
x_{0}\cos\theta+y_{0}\sin\theta-\sqrt{b^{2}\sin^{2}\theta+a^{2}\cos^{2}\theta} & \text{for the convex part}
\end{cases}
.\end{equation}

If we assume that $a = b = R$, which means that our ellipse is actually a circle, then equation~\ref{eq:lte_canon_form} matches the resulted equation of the previous work on the Legendre transform on circles~\cite{paper_legendre_circles}.

Now that we have found the Legendre transform of the ellipse in canonical form, we design and implement a method that utilizes the result of the transformation (equation~\ref{eq:lte_canon_form}) to find the common tangent of a set of ellipses.
The transformation of each ellipse of a set of ellipses in the Legendre space will result in a series of curves.
The intersections of these curves represent the common tangents of the transformed ellipses.
The point through which all of the curves pass is the common tangent of the set of ellipses.

\section{Description of the Method}
\label{sec:method_description}

In this section, we describe the method used for reconstructing tangent lines from a set of ellipses.

\subsection{Input}
\label{sec:model_description_first}

The input of the algorithm is a set of symmetrical ellipses and  the dimensions of the $\left( x, y \right) $ space that is to be analysed. An ellipse consists of it's center,  $\left(x_0,y_0\right)$, and it's major and minor axes, $a$ and $b$ respectively.

Based on the number of the given ellipses $n_{\mathrm{ellipse}}$, we calculate a maximum number of possible tangent lines $n_{\mathrm{lines}}$ as: $n_{\mathrm{lines}}=2 {n_{\mathrm{ellipse}}}/{n_{\mathrm{base}}} $ where $n_{\mathrm{base}}$ is a minimum expected number of ellipses per tangent and the coefficient $2$ means that the set of ellipses is expected to be symmetrical.
If that is the case, then there will two common tangent lines for each set of ellipses (figure~\ref{fig:ellipse_tangent_3s}).
Also, that number is always rounded up.
Although $n_{\mathrm{lines}}$ results in more tangent lines that should exist, during the execution of the algorithm the excess tangent lines will be discarded for various reasons we explain later.

\subsection{Clustering and extraction of lines}
\label{sec:cluster_extracion}

While transforming the ellipses to the Legendre space, we fill two-dimensional histograms with the results of equation~\ref{eq:lte_canon_form} for each given ellipse, iterating over $ \theta$.
The Legendre transform provides us with two curves per given ellipse.
Each curve raises the value of each bin through which it passes. Consequently, the intersections of the curves create peaks in the histograms.
Just like the intersections of the curves, the peaks represent the common tangent lines of the ellipses.
Since these curves are sinograms, they form clusters of high value bins around the peaks.
To detect the peaks in the histograms, the clusters that surround each peak must be extracted.
Therefore, the following clustering algorithm is  implemented.

\begin{figure}[h]\centering
         \includegraphics[width=0.66\linewidth]{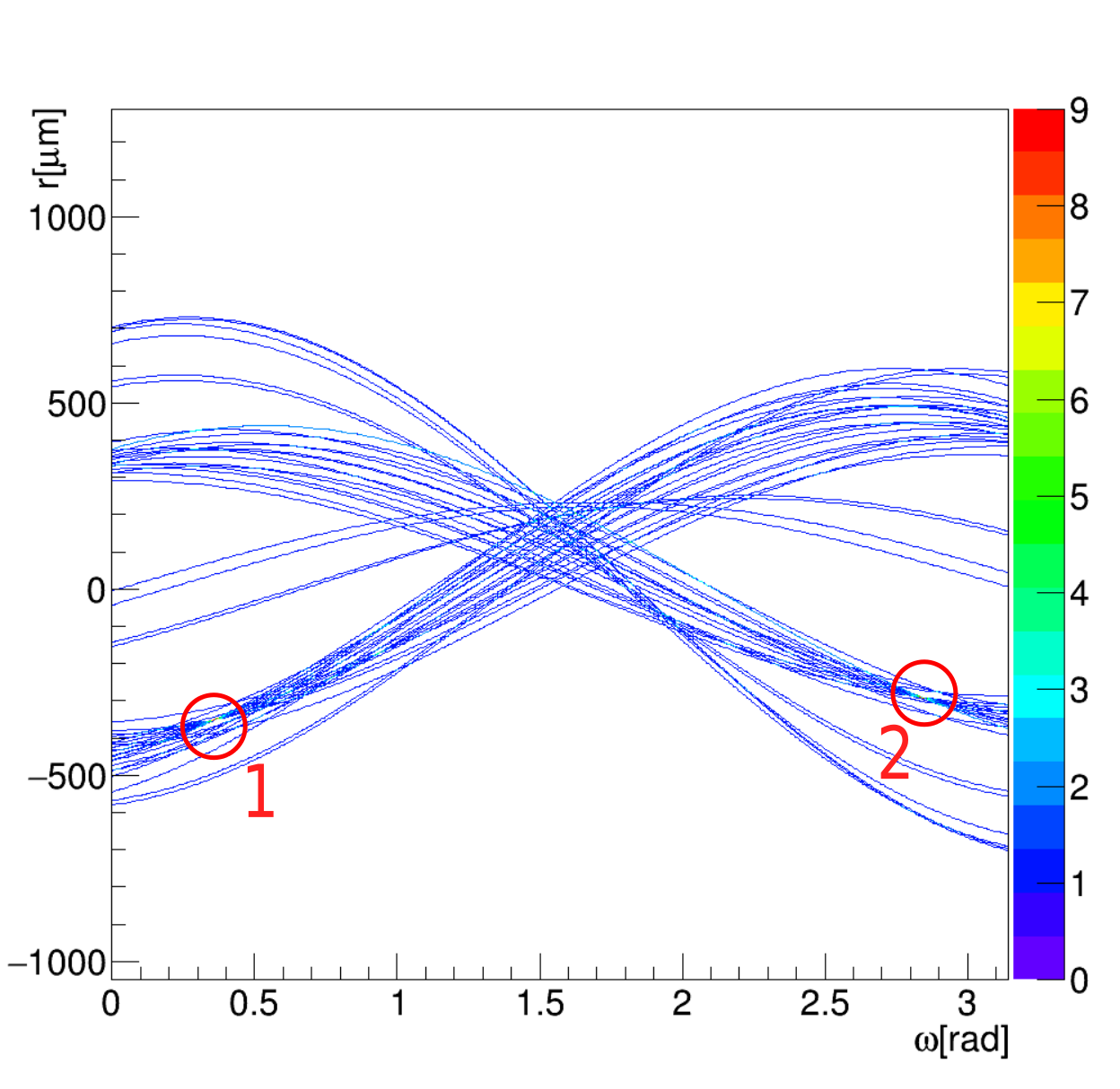}
         \caption{Example of a primary histogram with two peaks indicated with red circles.}
         \label{fig:hist_prim_ex}         
\end{figure}

\begin{figure}[h]\centering
    \begin{subfigure}[t]{0.75\textwidth}
         \includegraphics[width=0.98\linewidth]{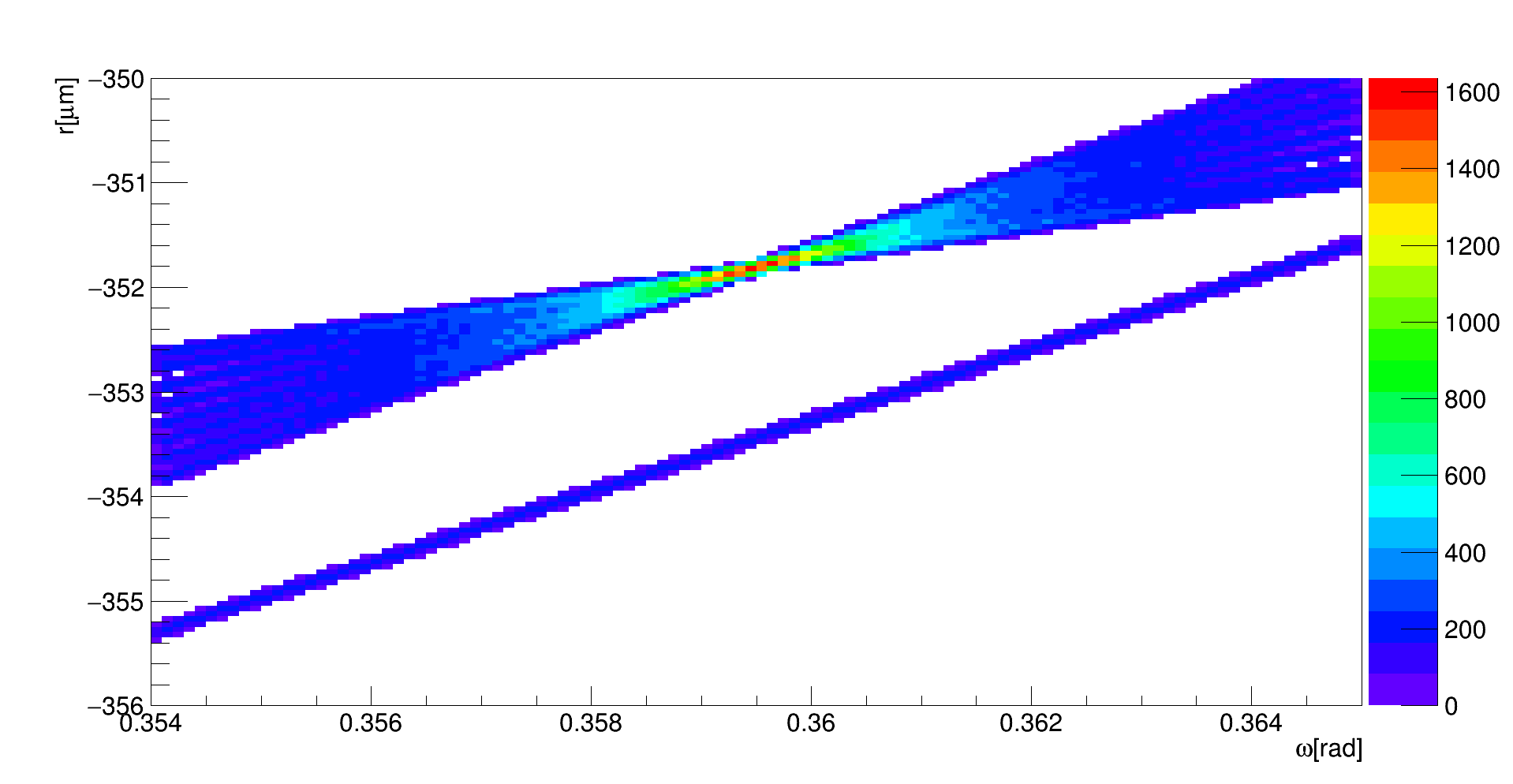}
         \caption{Secondary histogram for peak 1.}
	     \label{fig:hist_seco_ex_1}
    \end{subfigure}
    \\
    \begin{subfigure}[t]{0.75\textwidth}
         \includegraphics[width=0.98\linewidth]{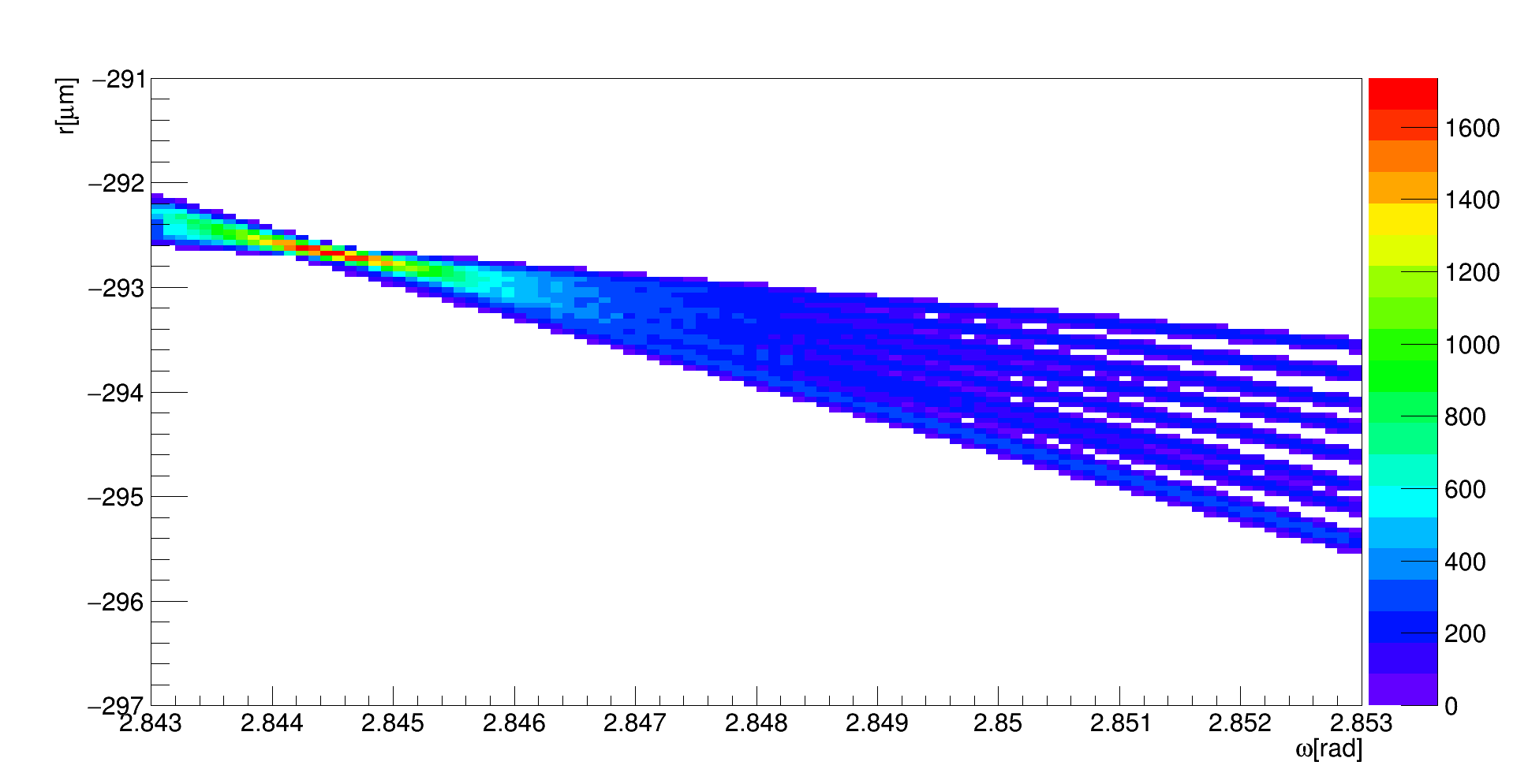}
         \caption{Secondary histogram for peak 2.}
	     \label{fig:hist_seco_ex_2}
    \end{subfigure}
    \caption{Examples of secondary histograms.}\label{fig:lte_histograms}
\end{figure}

The extraction of the tangent lines consists of two steps in which two dimensional histograms in the Legendre space are used.
First, a primary step during which we search for peaks in a single two-dimensional histogram (figure~\ref{fig:hist_prim_ex}) that covers all of the Legendre space that corresponds to the $\left( x, y \right) $ area that includes the ellipses.
Then, a secondary step, during which we review the areas with the highest peaks found in the primary histogram.
A two-dimensional histogram is used for each area (figures~\ref{fig:hist_seco_ex_1} and \ref{fig:hist_seco_ex_2}).


The histograms of the two steps differ from one another.
The first step uses a histogram that has a fixed number of bins for both axes and when a curve passes through a bin, the number of hits of that bin is increased by one.
This serves to reduce the load of the analysis and process quickly a large area whose most part does not give us any valuable information concerning the lines.
The second step uses fixed bin sizes for each axis and when a curve passes through a bin, a two-dimensional Gaussian signal is added on the histogram, using that bin as it's center.
This serves to either greatly increase the accuracy of the lines or reject the false positives.

\subsubsection{Primary step}
\label{sec:prim_step}

During the primary step, for the creation of the histogram,the histogram's size is addressed first.
We know that $\theta \in \left( 0, \pi \right) $ and the $r$-axis is calculated from equation~\ref{eq:lte_canon_form}.
To do so, the dimensions of the area, in the $\left(x,y\right)$ space, that is to be analysed is taken into account.
Therefore, we get:

$$
r\in\left(x_{\mathrm{min}}+y_{\mathrm{min}}-a_{\mathrm{max}},x_{\mathrm{max}}+y_{\mathrm{max}}+a_{\mathrm{max}}\right)
,$$
where $a_{\mathrm{max}}$ is the maximum value that the major axis $a$ of an ellipse can have. Due to the massive size of the $r$-axis, we use a fixed number of bins instead of fixed bin sizes.

\begin{figure}[h]\centering
    \begin{subfigure}[t]{0.4\textwidth}
         \includegraphics[scale=.50]{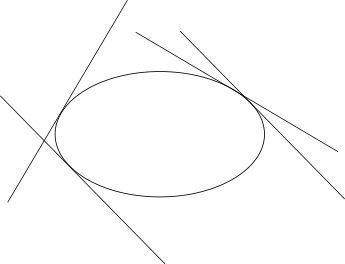}
         \caption{An ellipse has an infinite amount of tangents.}
         \label{fig:ellipse_tangent_1}
    \end{subfigure}
    ~
    \begin{subfigure}[t]{0.4\textwidth}
         \includegraphics[scale=.35]{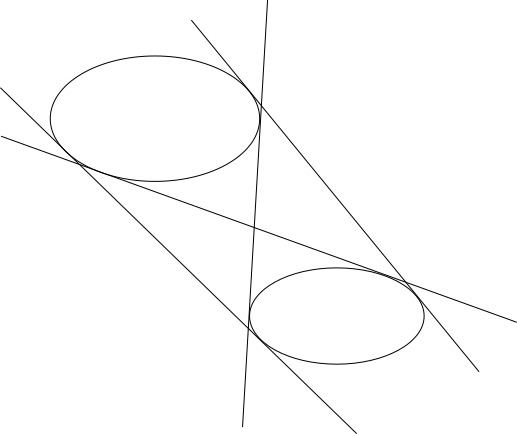}
         \caption{Two ellipses have four common tangents.}
	     \label{fig:ellipse_tangent_2}
    \end{subfigure}
    \\
    \begin{subfigure}[t]{0.4\textwidth}
         \includegraphics[scale=.40]{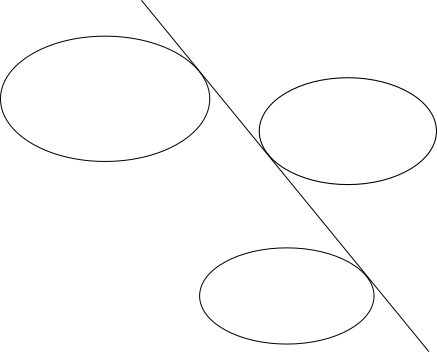}
         \caption{Three ellipses that have only one common tangent.}
         \label{fig:ellipse_tangent_3}
    \end{subfigure}
    ~
    \begin{subfigure}[t]{0.4\textwidth}
         \includegraphics[scale=.30]{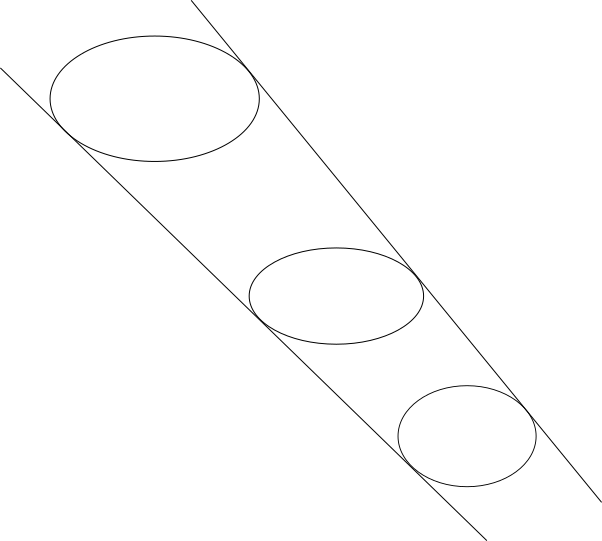}
         \caption{A symmetrical set of three ellipses that has two common tangents.}
	     \label{fig:ellipse_tangent_3s}
    \end{subfigure}

    \caption{The relation between ellipses through tangents for different sets of ellipses.}
\label{fig:ellipse_tangent}
\end{figure}

After filling the histogram, a threshold for the height of the bins of the histogram is applied and all bins under that specified threshold are ignored. 
The threshold is set to $3$ curves since a minimum of 3 ellipses in a set is required so that the set can have a maximum of two common tangent lines (figures~\ref{fig:ellipse_tangent_3} and \ref{fig:ellipse_tangent_3s}). Then, we sort the bins by height and then put them through an iterative process which attempts to result in clusters with the peaks at their centers. 

For each of the clusters, the bin with the greatest value is detected.
If the $70\% $ of its value is higher than the specified threshold, then, for that cluster only, the threshold is set to the  $70\% $ of the peak's value.
Then, we calculate the ratio of the sum of the hits (height) of each bin (above threshold) to the total number of bins (above threshold) of the cluster.
The smallest orthogonal boundary possible is set to each of the clusters according to the positions of the bins above threshold.
Finally, the mean values of $r$ and $\theta$ are  calculated, as well as the $\mathrm{slope}$ and the $\mathrm{intercept}$ with the corresponding errors:

$$
\begin{cases}
\delta\mathrm{slope}=\sqrt{\left(\frac{\partial \mathrm{slope}}{\partial\theta}\delta\theta\right)^{2}}\\
\delta\mathrm{intercept}=\sqrt{\left(\frac{\partial \mathrm{intercept}}{\partial\theta}\delta\theta\right)^{2}+\left(\frac{\partial \mathrm{intercept}}{\partial r}\delta r \right)^{2}}
\end{cases}
\Rightarrow
\begin{cases}
\delta\mathrm{slope}=\frac{1}{\sin^{2}\theta}\delta\theta\\
\delta\mathrm{intercept}=\frac{1}{\sin\theta}\sqrt{\left(r \cot\theta\delta\theta\right)^{2}+\left(\delta r \right)^{2}}
\end{cases}
.$$

The clusters are sorted based on their hits per bin ratios.
As possible solutions, we consider the $n_{\mathrm{lines}}$ clusters with the largest ratios.
These clusters will advance to the secondary step for further processing.

We should clarify that each cluster represents a different line and therefore a different solution to our problem.
This means that our algorithm always works for a set of ellipses that consists of  multiple subsets of ellipses, where each subset has it's own tangent, and assumes that the sets of ellipses are symmetric, meaning that two tangent lines correspond to each set, until proven otherwise.

\subsubsection{Secondary step}
\label{sec:seco_step}

During the secondary step, we create multiple two-dimensional histograms, one for each of the clusters with the greatest ratios from the primary histogram. The boundaries of each histogram is the boundary of the cluster being analysed.
This allows us to search each histogram for a single line instead of multiple lines as before.
In addition, due to the small size of the secondary histograms, we can use fixed bin sizes for both axes, $ \delta \theta $ and $\delta r $ for the $\theta$-axis and $r$-axis respectively.

As far as the $\theta$-axis is concerned, we choose the $ \delta \theta $ step for the bin size. On the other hand, the $r$-axis depends on $\theta$ and so the bin size will be calculated by applying the variance formula on equation~\ref{eq:lte_canon_form}:

$$
\delta r=\sqrt{\left(\frac{\partial r}{\partial a}\delta a\right)^{2}+\left(\frac{\partial r}{\partial b}\delta b\right)^{2}+\left(\frac{\partial r}{\partial\theta}\delta\theta\right)^{2}}
,$$
where $\delta a$ and $\delta b$ are the measurement errors of the major and minor axes of the ellipses respectively.
If we consider the worst error of both axes for both axes, meaning $\delta R = \max\left\lbrace \delta a, \delta b \right\rbrace $, we get:

$$
\delta r = \sqrt{\frac{\left(a\cos^{2}\theta\right)^{2}+\left(b\sin^{2}\theta\right)^{2}}{b^{2}\sin^{2}\theta+a^{2}\cos^{2}\theta}\delta R^{2}+\left(-x_{0}\sin\theta+y_{0}\cos\theta\pm\frac{\left(b^{2}-a^{2}\right)\sin\theta\cos\theta}{\sqrt{b^{2}\sin^{2}\theta+a^{2}\cos^{2}\theta}}\right)^{2}\delta\theta^{2}}
.$$
We also know that $ 0 < b \leq a $ and so $ b^2 \leq a^2 $, therefore:

$$
\delta r \leq \sqrt{\left(\cos^{4}\theta+\sin^{4}\theta\right)\delta R^{2}+\left(-x_{0}\sin\theta+y_{0}\cos\theta\right)^{2}\delta\theta^{2}}
.$$
Also, taking into account that $ \cos^{4}\theta+\sin^{4}\theta \leq 1$, we conclude that:

$$
\delta r\leq\sqrt{\delta R^{2}+\left(-x_{0}\sin\theta+y_{0}\cos\theta\right)^{2}\delta\theta^{2}}
.$$
Therefore, the error of the line parameter $r$ consists of the angle step $\delta \theta $, the expected measurement error $\delta R$ and the center of the ellipse $ \left(x_{0}, y_{0} \right) $.
To ensure consistency in our results, we prefer a uniform binning, independent of the bin under examination and the position of the ellipse.
For this purpose, we select the angle step  $ \delta \theta $ to be small enough so that $\delta r \gg \delta \theta \Rightarrow \delta r \approx \delta R $.

To have the greatest accuracy possible, the sizes of the bins are going to have to be as small as possible. That may cause errors in noisy environments since the hits will not create a peak in a certain bin, but rather spread around. Therefore, to keep the steps small but avoid errors, we turn to an alternative solution.

The key for creating peaks in a histogram that otherwise, due to errors, would not exist, is the use of a Gaussian kernel based analysis. It has been shown that the use of Gaussian sums in histograms increases the effectiveness of algorithms and makes them robust against noise \cite{paper_gaussian_sums}. So,  we apply Gaussian signals instead of simply increasing the hits of the appropriate bins during the transformation to the Legendre space.

The mean of a Gaussian signal will be the center of the $\theta$-axis bin being iterated and the corresponding $r \left( \theta \right)$ value, and the steps $\delta \theta$ and $\delta r$ will serve as the standard deviations $ \sigma_{\theta}$ and $ \sigma_{r}$ respectively, concluding to a Gaussian signal:
\begin{equation}
\label{eq:gauss_signal}
f_{\mathrm{gauss}}(\theta,r)=\mathrm{A}\exp\left(-\left(\frac{(\theta-\theta_{0})^{2}}{2\sigma_{\theta}^{2}}+\frac{(r-r_{0})^{2}}{2\sigma_{r}^{2}}\right)\right)
,\end{equation}
where $\mathrm{A}$ is an arbitrary constant.

Therefore, for every $\theta$-axis bin in the histogram, Gaussian signals (equation~\ref{eq:gauss_signal}) for each part of each ellipse are  calculated and added to the histogram.
The curves for the convex and the concave part of an ellipse in the Legendre space should never intersect with each other\footnote{That would imply that there is a common tangent line for both the concave and the convex part of the ellipse.}  but we still have to evaluate both of them since we do not have a priori knowledge of which of the curves pass through the cluster under consideration.
The same goes for all the ellipses as well, meaning that not all of them will have a tangent line represented by the specific $\left( \theta, r \right)$ pair in the histogram under construction.

To maintain a low computational cost, the Gaussian signal is truncated. In other words, the value of the signal is added to the histogram only if it is above a specified threshold. Therefore, the iterative process starts from the peak of the Gaussian signal and continues outwards in consecutive boxes until it reaches the aforementioned threshold where the iterative processes stops.

Apart for the aforementioned differences, the rest of the procedure is same as in the primary step. We apply the thresholds, narrow down the peak of each cluster, sort the clusters based on their hits per bin ratios, calculate the mean values, and extract the line parameters.

\subsection{Improving the lines' characteristics}
\label{improve_line}

After tracking the lines, we can further increase the accuracy of their characteristics or dismiss false positives with a $3$ step procedure.

\subsubsection{Line-Ellipse proximity test}
\label{sec:proximity}

As a first step, if possible, the given ellipses are associated with the detected lines if they fulfil a certain criterion: The distance $d$ between the detected line and the closest point of the ellipse to the line must be smaller than $n_{\sigma}$ times the value of $\sigma_{r}$:
$$
d_{\mathrm{closest}} < n_{\sigma}  \sigma_{r}
,$$
where $n_{\sigma}$ is an arbitrary constant integer.

\begin{figure}[h]\centering
         \includegraphics[scale=.42]{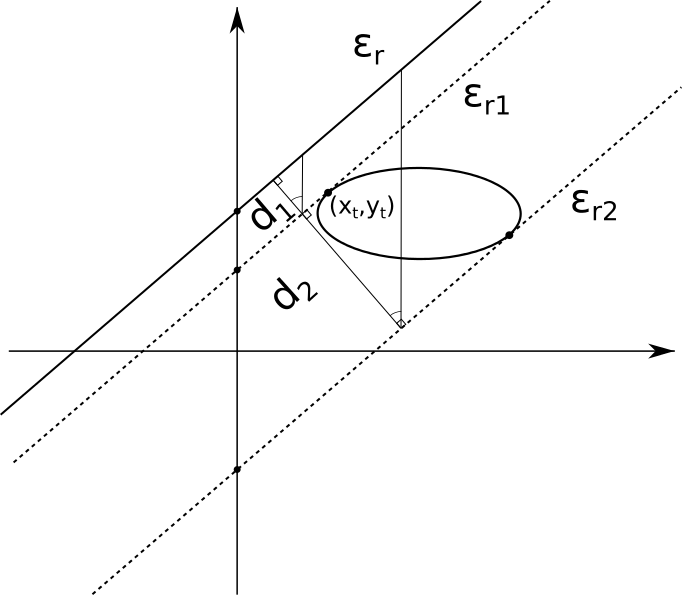}
        \caption{Detected line $\varepsilon_{r}$ and the parallels $\varepsilon_{r1}$ and $\varepsilon_{r2}$. In this case, $\varepsilon_{r1}$ will be chosen.}
	\label{fig:tracks_parallel_ellipse}
\end{figure}

For each ellipse, we create two lines parallel to our detected line as in figure~\ref{fig:tracks_parallel_ellipse}.
One parallel line tangent to the concave part of the ellipse and the other parallel line tangent to the convex part of the ellipse. We keep the line closest to the detected line and then apply the criterion.

If a line has less ellipses that a required minimum, we consider it to be a false positive and therefore reject it.

\subsubsection{Least square fit}
\label{sec:linear_fit}

For each line with an adequate number of ellipses, we can apply a linear fit to increase their accuracy.
Prior to that, a series of tangent points must be located between the lines and their associated ellipses.

To locate the tangent point between a line and its ellipse, the closest parallel is used (figure~\ref{fig:tracks_parallel_ellipse}). Therefore, for a parallel line of the form $y = \mathrm{slope} \cdot x + \mathrm{intercept} $, the tangent points $ \left( x_{t},y_{t} \right)$ are calculated using the equations:
$$
x_{t}=x_{0}-\frac{\mathrm{slope}\cdot a^{2}}{\left(\mathrm{intercept}+\mathrm{slope}\cdot x_{0}-y_{0}\right)}
$$
$$
y_{t}=y_{0}+\frac{b^{2}}{\mathrm{intercept}+\mathrm{slope}\cdot x_{0}-y_{0}}
,$$
where $ \left(x_{0}, y_{0} \right) $ is the center of the ellipse and $a$, $b$ the ellipse's major and minor axis respectively.

We repeat this task for each of the ellipses associated with a detected line. Then, a linear fit is performed on the tangent points. We use the attributes of the detected line as initial values for the fitting procedure.

\subsubsection{Chi-square test}
\label{sec:chi_square}

During the final stage of the algorithm, we perform a $\chi^2$ test on the fitted lines.
The ratio of $\chi^2$ to the Number of Degrees of Freedom (NDF) \footnote{In the final calculation of a statistic, the number of values that are free to vary is the number of degrees of freedom.} is required to be less than $0.5$ 
$\left( {\chi^{2}}/{\mathrm{NDF}}<0.5 \right)$.
Lines that don't meet the requirement are rejected.
This prevents from noise induced ellipses to create prevailing lines.

\section{Performance Studies}
\label{sec:application}

\begin{figure}[h]\centering
         \includegraphics[scale=.45]{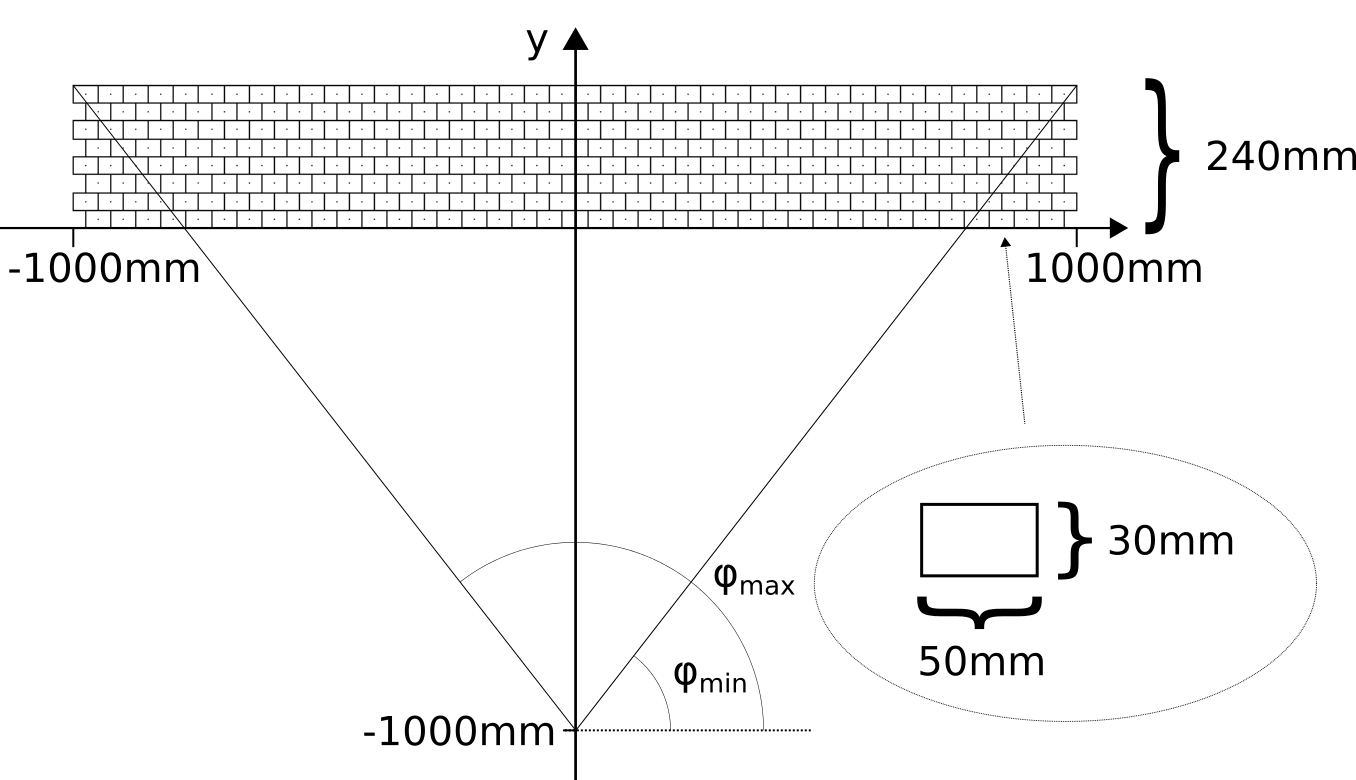}
        \caption{Specifications of the detector and the chambers.}
	\label{fig:det_specs}
\end{figure}

To study our method, we implement a hypothetical detector as illustrated in figure~\ref{fig:det_specs}.
The detector consists of layers of chambers, the intersection of which is orthogonal.
Using toy Monte Carlo generated lines that represent the tracks of a hypothetical charged particle, ellipses in the appropriate chambers are  created.
Then, we apply the smearing of the ellipse's characteristics and the noise of the electronics. Our proposed method reconstructs the tracks based on these ellipses.
Finally, we compare the attributes of the initial, Monte Carlo generated, tracks with the attributes of the reconstructed tracks.

\subsection{Description of the hypothetical detector}
\label{sec:det_specs}

The hypothetical detector consists of $8$ layers of chambers (table \ref{tab:det_specs}).
These chambers have an orthogonal intersection $50\,\mathrm{mm}$ wide and $30\,\mathrm{mm}$ high.
The detector spans from $-1 \mathrm{m}$ to $+1 \mathrm{m}$ on $x$-axis and it's base is placed at $y=0$.
The height of the detector depends on the number of layers and the height of the chambers.
Therefore, our detector is $240\,\mathrm{mm}$ high.

\begin{table}[h!]
\begin{center}
\caption{\normalsize Detectors' specifications.}
\vspace{1ex}
\begin{tabular}{c c}
\hline\hline
Attribute & Value \\\hline
{Number of layers} $\left(n_\mathrm{layers}\right)$  & $ 8$  \\
$ x_\mathrm{min} $  & $ -1\,\mathrm{m}$  \\
$ x_\mathrm{max} $  & $ +1\,\mathrm{m}$  \\
{Base} $\left(y_\mathrm{min}\right) $  & $ 0\,\mathrm{mm}$  \\
{Top} $\left( y_\mathrm{max}\right) $  & $ 240\,\mathrm{mm}$  \\
{Chamber} {width}   & $ 50\,\mathrm{mm}$  \\
{Chamber} {height}  & $ 30\,\mathrm{mm}$  \\\hline\hline
\end{tabular}
\label{tab:det_specs}
\end{center}
\end{table}

Since the detector is purely hypothetical, we do not take into account technical characteristics such as the thickness of the chambers boundaries. Instead, we focus on the mathematical approach of our pattern recognition method. This does not cause any problems because our method is generic and independent of the problem's conditions, which means that it does not need such details to operate.

\subsection{Toy Monte Carlo}
\label{sec:mc}

We implement a Monte Carlo algorithm that produces random tracks.
These tracks pass through the detector and in each of the chambers through which a track passes, ellipses that are co-eccentric with the chambers are created.
The track must be tangent to the ellipses of the chambers through which it passed.

A noisy environment, such as a measurement error or the noise of the electronics, can also be simulated.
The first can be simulated by the smearing of the characteristics of the ellipse's major and minor axes.
We implement this by applying a Gaussian error on the aforementioned characteristics.
The noise of the electronics can be simulated by generating random ellipses in empty chambers.

Finally, the ellipses are given as input to our method and the resulted tracks are returned.
To perform a quantitative evaluation of our method, we introduce the suitable line parameters $ \mathrm{slope} = \tan\phi$ and the $\mathrm{intercept}$ with the $y$-axis.
The efficiency of the algorithm is based on the difference between the parameters of the initial tracks and the parameters of the reconstructed tracks.

\subsubsection{Reconstruction efficiency, fake rate, and fail rate}
\label{sec:efficiency}

To evaluate a reconstructed track in comparison to it's initial, Monte Carlo generated, track, we address the relative errors.
The relative errors of a reconstructed track of the form $ y = \mathrm{slope_{reco}} x + \mathrm{intercept_{reco}} $ to an initial track of the form $ y = \mathrm{slope_{init}} x + \mathrm{intercept_{init}} $  are calculated as:
$$
\mbox{Relative error of the slope} = \frac{\mathrm{slope_{init}} - \mathrm{slope_{reco}}}{ \mathrm{slope_{init}}} 100 \%
$$
and
$$
\mbox{Relative error of the intecept} = \frac{\mathrm{intecept_{init}} - \mathrm{intecept_{reco}}}{ \mathrm{intecept_{init}}} 100 \%
.$$

To facilitate the evaluation of the performance of our method, we introduce the reconstruction efficiency.
The reconstruction efficiency, for which we will refer to as efficiency from now on, is the ratio of matched tracks $\left( N_{\mathrm{match}} \right)$ to initial tracks $\left( N_{\mathrm{init}} \right)$.
A reconstructed track is considered to match an initial track if the absolute value of the relative error of the slope and the absolute value of the relative error of the intercept and  are less than the specified thresholds $\mathrm{slope_{threshold}}$ and $\mathrm{intercept_{threshold}}$ respectively. 

If the above criteria are not fulfilled, the track is classified as fake. We also define the fake rate as the ratio of reconstructed tracks that were not matched to an initial track, $N_{\mathrm{fake}}$, to initial tracks.

The fail rate is defined as lack of results in comparison to the number of results expected.
This means that every time the algorithm fails to return a number of tracks, equal or more than the initial number of tracks, the fail counter rises by the number of tracks it failed to produce.

\subsubsection{Variables, constants, and thresholds}
\label{sec:variables}

As we have already mentioned, a number of specified variables, constants, and thresholds is used.
We will mention some of them to give an even better insight to our analysis.

First of all, during the primary step of the Legendre based analysis, we set the bin numbers for the two dimensional histograms at $1000$ bins for both axes.
On the contrary, during the secondary step, we set the size of the bins for the two dimensional histograms at $ \delta_{\theta} = 10^{-4}\,\mathrm{rad}$  and $\delta_{r} = 0.05\,\mathrm{\upmu m}$ for the $\theta$ and $r$ axes respectively. The bin sizes are also used as standard deviations for the Gaussian signals,  $ \sigma_{\theta} = \delta_{\theta}$  and $\sigma_{r} = \delta_{r} $.

The coefficient $n_{\sigma}$ for the line-ellipse proximity test is set to $20$. This means that the criterion for the maximum distance between a line and an ellipse becomes: $d_{\mathrm{closest}} < n_{\sigma}  \sigma_{r} = 1\,\mathrm{mm} $.

Finally, to evaluate the efficiency of the algorithm, both the $\mathrm{slope_{threshold}}$ and the $\mathrm{intercept_{threshold}}$ are set to $1\%$. This means that only reconstructed tracks whose characteristics differ less than $1\%$ from their initial, Monte Carlo generated, tracks are accepted as valid results. The rest of the reconstructed tracks are classified as fake.

\subsection{Resolution}
\label{reso}

An event is defined as the generation of $n$ Monte Carlo tracks, the creation of the ellipses based on the aforementioned tracks, and the reconstruction of the tracks through the use of our method. 
A Monte Carlo generated track intercepts the $y$-axis  $1\,\mathrm{m}$ below the detector and has a random angle $\phi$.
We evaluate our method for single and multi-track events.

\subsubsection{Single track events}
\label{sec:single_track}

Single track events are generated for different noise parameters.
First, for different values of smearing, but a steady noise level at $0\%$.
Then, for different noise levels and the smearing $=0\,\mathrm{\upmu m} $.

\begin{figure}[h]\centering

    \begin{subfigure}[t]{0.48\textwidth}
         \includegraphics[scale=.4]{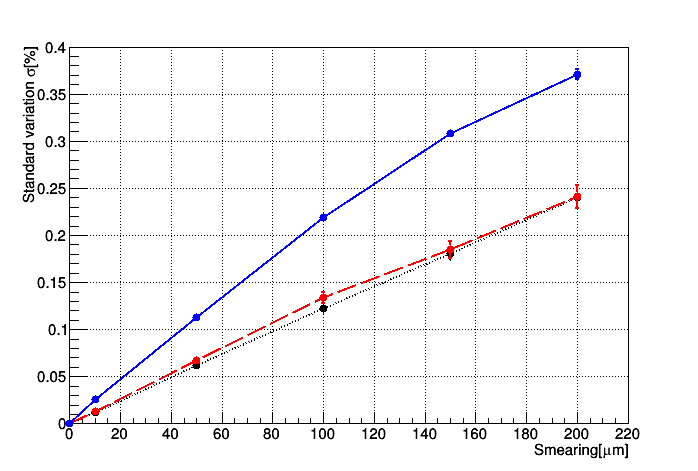}
         \caption{Standard variation of the relative errors slope (red-dashed line), intercept (black-dotted line), and of the residuals (blue line).}
	     \label{fig:std_var}
    \end{subfigure}
    ~
    \begin{subfigure}[t]{0.48\textwidth}
         \includegraphics[scale=.4]{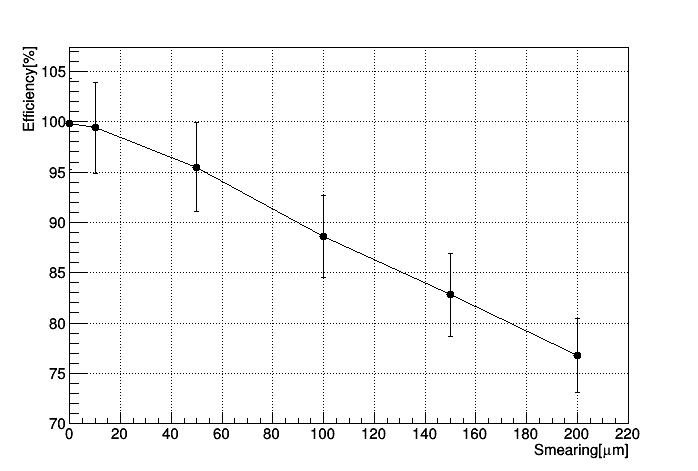}
         \caption{Efficiency of the algorithm versus different values of smearing.}
         \label{fig:efficiency}
    \end{subfigure}
    \caption{Results with respect to smearing. Noise level at $0\%$. }
\label{fig:graph_smearings}
\end{figure}

All the results for a steady noise level of $0\%$ and different values of smearing, ranging from $0\,\mathrm{\upmu m}$ to $200\,\mathrm{\upmu m}$, are presented in table \ref{tab:rates_smearing}.
Despite the sudden rise of the standard variations from a smearing of $0\,\mathrm{\upmu m}$ to $10\,\mathrm{\upmu m}$, there is a linear relation between the standard variations of all slope, intercept, and residuals with respect to smearing (figure~\ref{fig:std_var}).
In addition, the efficiency also declines linearly, as the values of smearing increases (figure~\ref{fig:efficiency}).

\begin{table}[h!]
\begin{center}
\caption{\normalsize Efficiency of the algorithm for different values of $ \mathrm{smearing}$.}
\label{tab:rates_smearing}

\begin{subtable}{.99\textwidth}
\begin{center}
\vspace{1ex}
\caption{\normalsize Values of $ \mathrm{smearing}$ from $ 0 \mbox{ to } 50\,\mathrm{\upmu m} $.}
\vspace{1ex}
\begin{tabular}{c c c c}
\hline\hline
Smearing & $0\,\mathrm{\upmu m}$ & $10\,\mathrm{\upmu m}$ & $50\,\mathrm{\upmu m}$  \\\hline
Efficiency [\%] &
 $ 99.8 \pm 4.5 $ & $ 99.4 \pm 4.5 $ & $ 95.5 \pm 4.4 $ \\ 
Fake rate  [\%] &
 $  0.0 \pm 0.0 $ & $  0.0 \pm 0.0 $ & $  0.6 \pm 0.3 $ \\
Fail rate  [\%] &
 $  0.2 \pm 0.1 $ & $  0.6 \pm 0.3 $ & $  3.9 \pm 0.6 $ \\
 \hline\hline
 $\sigma_\mathrm{slope}$ [\%] &
 $ \left(1.04 \pm 0.04 \right) \times 10^{-5} $&
 $ \left(1.21 \pm 0.05 \right) \times 10^{-2} $&
 $ \left(6.16 \pm 0.26 \right) \times 10^{-2} $\\
 $\sigma_\mathrm{intercept}$ [\%] &
 $ \left(1.08 \pm 0.04 \right) \times 10^{-5} $&
 $ \left(1.36 \pm 0.05 \right) \times 10^{-2} $&
 $ \left(6.74 \pm 0.28 \right) \times 10^{-2} $\\
 $\sigma_\mathrm{residuals}$ [$\mathrm{\upmu m}$] &
 $ \left(2.55 \pm 0.03 \right) \times 10^{-4} $&
 $ \left(2.56 \pm 0.03 \right) \times 10^{-2} $&
 $ \left(1.13 \pm 0.02 \right) \times 10^{-1} $ \\
 \hline\hline
\end{tabular}
\end{center}
\end{subtable}

\begin{subtable}{.99\textwidth}
\begin{center}
\vspace{1ex}
\caption{\normalsize Values of $ \mathrm{smearing}$ from $ 100 \mbox{ to } 200\,\mathrm{\upmu m} $.}
\vspace{1ex}
\begin{tabular}{c c c c }
\hline\hline
Smearing & $100\,\mathrm{\upmu m}$ & $150\,\mathrm{\upmu m}$ & $200\,\mathrm{\upmu m}$ \\\hline
Efficiency [\%] &
 $ 88.6 \pm 4.1 $ & $ 82.8 \pm 4.1 $ & $ 76.8 \pm 3.7   $ \\ 
Fake rate  [\%] &
 $  2.7 \pm 0.5 $ & $  4.4 \pm 0.5 $ & $  4.9 \pm 0.7   $ \\
Fail rate  [\%] &
 $  8.7 \pm 1.0 $ & $ 13.8 \pm 1.0 $ & $ 18.3 \pm 1.5   $ \\
 \hline\hline
 $\sigma_\mathrm{slope}$ [\%] &
 $ \left(1.22 \pm 0.05 \right) \times 10^{-1} $&
 $ \left(1.81 \pm 0.08 \right) \times 10^{-1} $&
 $ \left(2.40 \pm 0.12 \right) \times 10^{-1} $ \\
 $\sigma_\mathrm{intercept}$ [\%] &
 $ \left(1.34 \pm 0.06 \right) \times 10^{-1} $&
 $ \left(1.85 \pm 0.09 \right) \times 10^{-1} $&
 $ \left(2.41 \pm 0.12 \right) \times 10^{-1} $ \\
 $\sigma_\mathrm{residuals}$ [$\mathrm{\upmu m}$] &
 $ \left(2.19 \pm 0.03 \right) \times 10^{-1} $&
 $ \left(3.08 \pm 0.04 \right) \times 10^{-1} $&
 $ \left(3.71 \pm 0.05 \right) \times 10^{-1} $ \\
 \hline\hline
\end{tabular}
\end{center}
\end{subtable}

\end{center}
\end{table}

\begin{figure}[h]\centering
    \begin{subfigure}[t]{0.4\textwidth}
         \includegraphics[scale=.3]{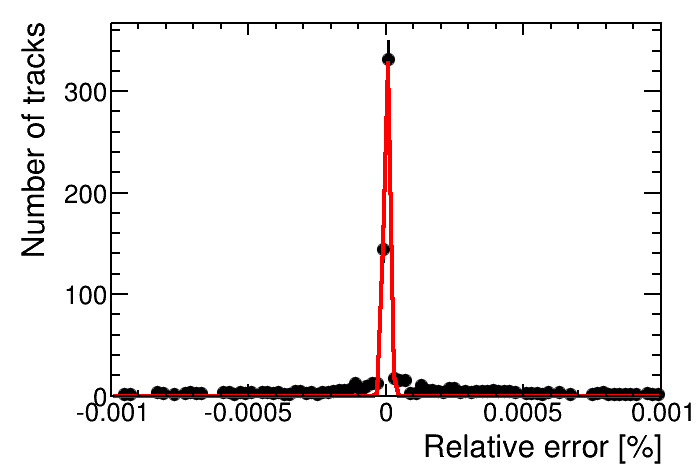}
         \caption{Relative errors of slope [\%]}
         \label{fig:lte_mc_s0000_n0000_mc_s}
    \end{subfigure}
    ~
    \begin{subfigure}[t]{0.4\textwidth}
         \includegraphics[scale=.3]{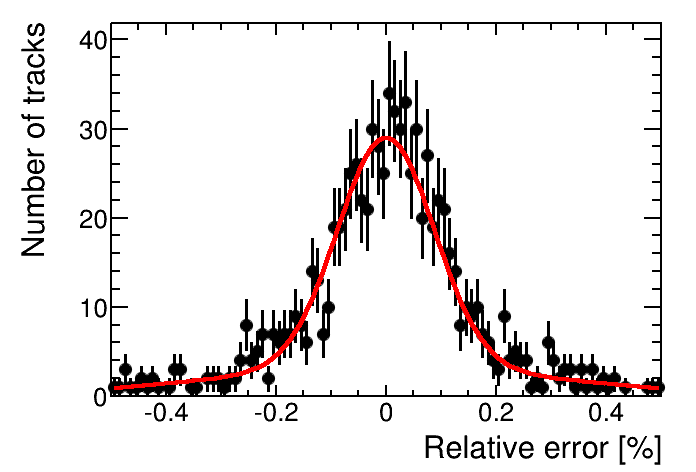}
         \caption{Relative errors of slope [\%]}
         \label{fig:lte_mc_s0100_n0000_mc_s}
    \end{subfigure}
    \\
    \begin{subfigure}[t]{0.4\textwidth}
         \includegraphics[scale=.3]{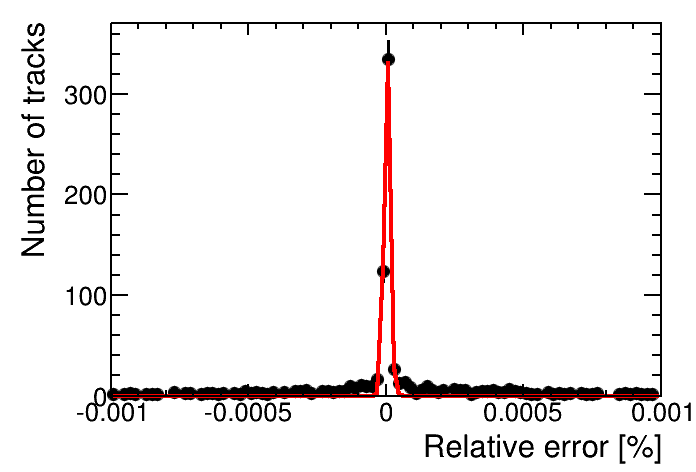}
         \caption{Relative errors of intercept [\%]}
	     \label{fig:lte_mc_s0000_n0000_mc_i}
    \end{subfigure}
    ~
    \begin{subfigure}[t]{0.4\textwidth}
         \includegraphics[scale=.3]{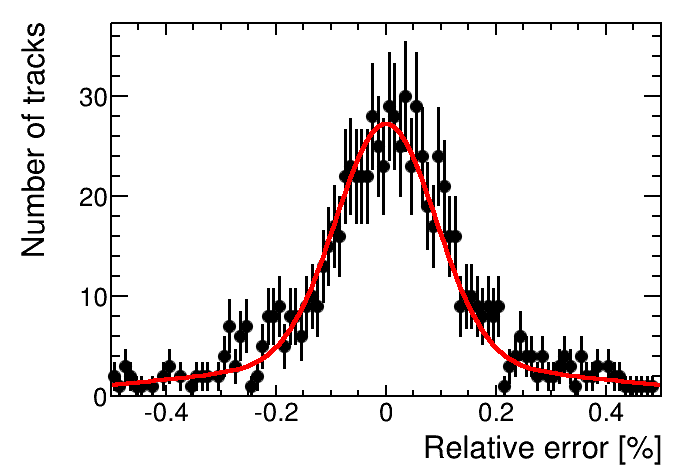}
         \caption{Relative errors of intercept [\%]}
	     \label{fig:lte_mc_s0100_n0000_mc_i}
    \end{subfigure}
    \\
    \begin{subfigure}[t]{0.4\textwidth}
         \includegraphics[scale=.3]{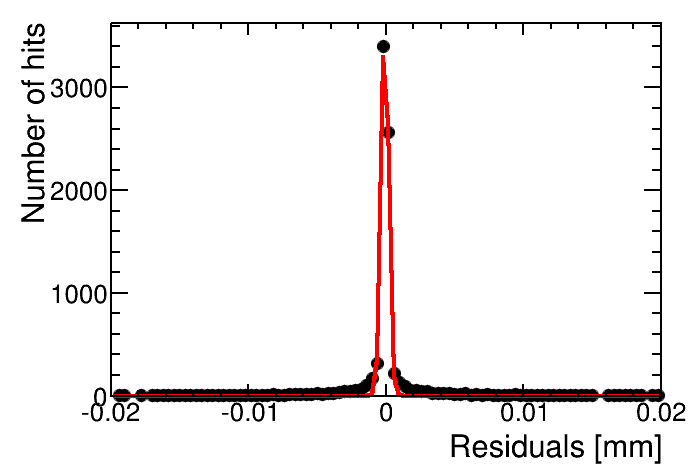}
         \caption{Residuals [$\mathrm{mm} $]}
	     \label{fig:resi_mc_s0000_n0000_mc}
    \end{subfigure}
    ~
    \begin{subfigure}[t]{0.4\textwidth}
         \includegraphics[scale=.3]{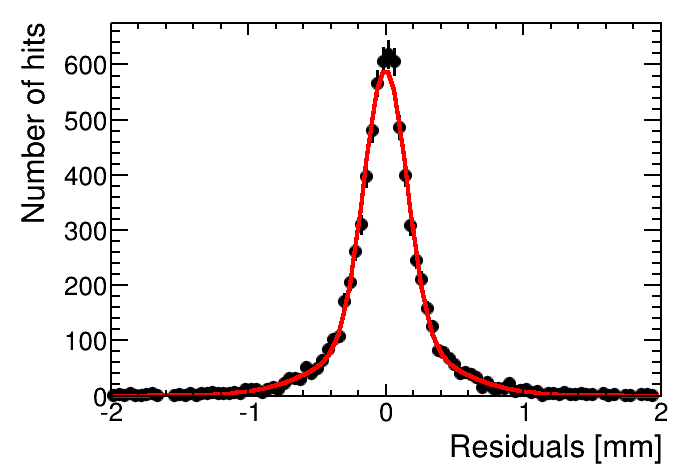}
         \caption{Residuals [$\mathrm{mm} $]}
	     \label{fig:resi_mc_s0100_n0000_mc}
    \end{subfigure}

    \caption{Histograms of the Monte Carlo results with $0 \% $ $ \mathrm{ noise}$. The left column has the results with $\mathrm{smearing} = 0\,\mathrm{\upmu m} $ while the right column with $ \mathrm{smearing} = 0.1\,\mathrm{\upmu m} $. The lines, from top to bottom, illustrate the relative errors of $\mathrm{slope}$, the relative errors of the $\mathrm{intercept} $ with the $y$-axis, and the residuals. }
\label{fig:lte_smearing}
\end{figure}

At figure~\ref{fig:lte_smearing} we illustrate the results of the simulations for $ \mathrm{smearing} = 0\,\mathrm{\upmu m} $  (left column) and $ \mathrm{smearing} = 0.1\,\mathrm{\upmu m} $ (right column) but $0\%$ noise in both cases. 
The relative errors of the slope (figures~\ref{fig:lte_mc_s0000_n0000_mc_s} and \ref{fig:lte_mc_s0100_n0000_mc_s}), the intercept with the $y$-axis(figures~\ref{fig:lte_mc_s0000_n0000_mc_i} and \ref{fig:lte_mc_s0100_n0000_mc_i}), and the residuals (figures~\ref{fig:resi_mc_s0000_n0000_mc} and \ref{fig:resi_mc_s0100_n0000_mc}) of each of the reconstructed tracks of the Monte Carlo experiment in comparison to their initial tracks have Gaussian distributions.

All the results for  different noise levels, ranging from $0\%$ to  $150\%$ and a steady value of smearing at $0\,\mathrm{\upmu m}$, are presented in table \ref{tab:rates_noise}. The efficiency of the algorithm is always above 98\%, even with the noise at the high level of $ 150\%$.

\begin{table}[h!]
\begin{center}
\caption{\normalsize Efficiency of the algorithm for different values of $ \mathrm{noise}$.}
\label{tab:rates_noise}

\begin{subtable}{.99\textwidth}
\begin{center}
\vspace{1ex}
\caption{\normalsize Values of $ \mathrm{noise}$ from $ 0\% \mbox{ to } 50\%$.}
\vspace{1ex}

\begin{tabular}{c c c c}
\hline\hline
Noise level & $  0\,\%$& $ 10\,\%$& $ 50\,\%$ \\\hline
Efficiency [\%] & 
 $  99.8 \pm 4.5 $ &
 $  99.5 \pm 4.5 $ &
 $  98.8 \pm 4.5 $ \\  
Fake rate  [\%] & 
 $   0.0 \pm 0.0 $ &
 $   0.1 \pm 0.1 $ &
 $   0.6 \pm 0.3 $ \\
Fail rate  [\%] & 
 $   0.2 \pm 0.1 $ &
 $   0.4 \pm 0.2 $ &
 $   0.5 \pm 0.2 $ \\
 \hline\hline
 $\sigma_\mathrm{slope}$ [\%] &
 $ \left(1.04 \pm 0.04 \right) \times 10^{-5} $&
 $ \left(1.14 \pm 0.04 \right) \times 10^{-5} $&
 $ \left(1.08 \pm 0.04 \right) \times 10^{-5} $\\
 $\sigma_\mathrm{intercept}$ [\%] &
 $ \left(1.08 \pm 0.04 \right) \times 10^{-5} $& 
 $ \left(1.12 \pm 0.04 \right) \times 10^{-5} $&
 $ \left(1.11 \pm 0.04 \right) \times 10^{-5} $\\
 $\sigma_\mathrm{residuals}$ [$\mathrm{\upmu m}$] &
 $ \left(2.55 \pm 0.03 \right) \times 10^{-4} $&
 $ \left(2.62 \pm 0.03 \right) \times 10^{-4} $&
 $ \left(2.55 \pm 0.03 \right) \times 10^{-5} $\\
 \hline\hline
\end{tabular}
\end{center}
\end{subtable}

\begin{subtable}{.99\textwidth}
\begin{center}
\vspace{1ex}
\caption{\normalsize Values of $ \mathrm{noise}$ from $ 100\% \mbox{ to } 150\%$.}
\vspace{1ex}
\begin{tabular}{c c c}

\hline\hline
Noise level & $100\,\%$ & $150\,\%$ \\\hline
Efficiency [\%] & 
 $  99.2 \pm 4.5 $ &
 $  98.2 \pm 4.5 $ \\  
Fake rate  [\%] & 
 $   0.4 \pm 0.2 $ &
 $   0.7 \pm 0.2  $ \\
Fail rate  [\%] & 
 $   0.4 \pm 0.2 $ &
 $   1.1 \pm 0.3 $ \\
 \hline\hline
 $\sigma_\mathrm{slope}$ [\%] &
 $ \left(1.14 \pm 0.04 \right) \times 10^{-5} $&
 $ \left(1.14 \pm 0.04 \right) \times 10^{-5} $ \\
 $\sigma_\mathrm{intercept}$ [\%] &
 $ \left(1.10 \pm 0.04 \right) \times 10^{-5} $&
 $ \left(1.07 \pm 0.03 \right) \times 10^{-5} $ \\
 $\sigma_\mathrm{residuals}$ [$\mathrm{\upmu m}$] &
 $ \left(2.37 \pm 0.03 \right) \times 10^{-5} $&
 $ \left(2.20 \pm 0.02 \right) \times 10^{-5} $ \\
 \hline\hline

\end{tabular}
\end{center}
\end{subtable}

\end{center}
\end{table}

\subsubsection{Multi-track events}
\label{sec:multi_track}

The method is also checked with multi-track events (table \ref{tab:multi_track_results}).
The number of tracks barely affects the efficiency of algorithm and that the quality of the the reconstructed tracks' characteristics has been preserved.

In figure~\ref{fig:dual_track_ex}, an example of a dual-track event with $50\%  \mbox{ noise}$, but without any $ \mathrm{smearing}$ at all, is illustrated.
Both tracks were reconstructed successfully even though the left track has noise induced ellipses in close proximity. In a similar fashion, in figure~\ref{fig:dual_track_ex_robust}, there is a similar example of a dual with the same conditions as before. This time, both tracks have been reconstructed successfully, even though they are too close to each other and at same time, almost parallel. Despite the low noise level, in the area that contains the tracks' ellipses, we have a great ellipse density. This means that the ellipses of the left track, are noise for the ellipses of the right track and vice versa.

\begin{landscape}
\begin{figure}[h]\centering
         \includegraphics[scale=.42]{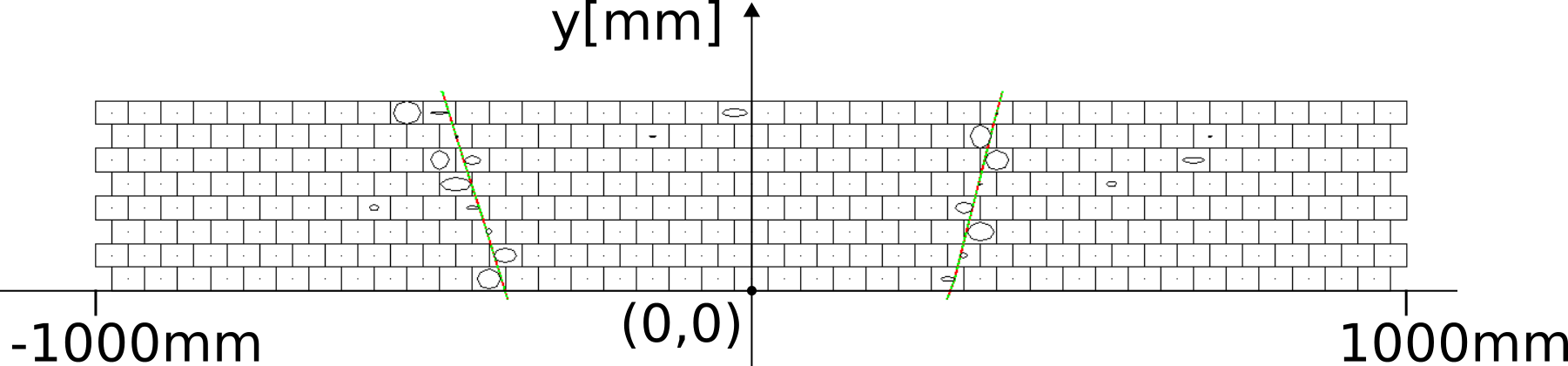}
        \caption{A dual track event with $50\%$ noise. The green lines are the reconstructed tracks. The red dashed lines are the initial lines for reference. From this set of ellipses, the algorithm created the histograms in figures~\ref{fig:hist_prim_ex} and \ref{fig:lte_histograms}.}
	\label{fig:dual_track_ex}
\end{figure}

\begin{figure}[h]\centering
         \includegraphics[scale=.41]{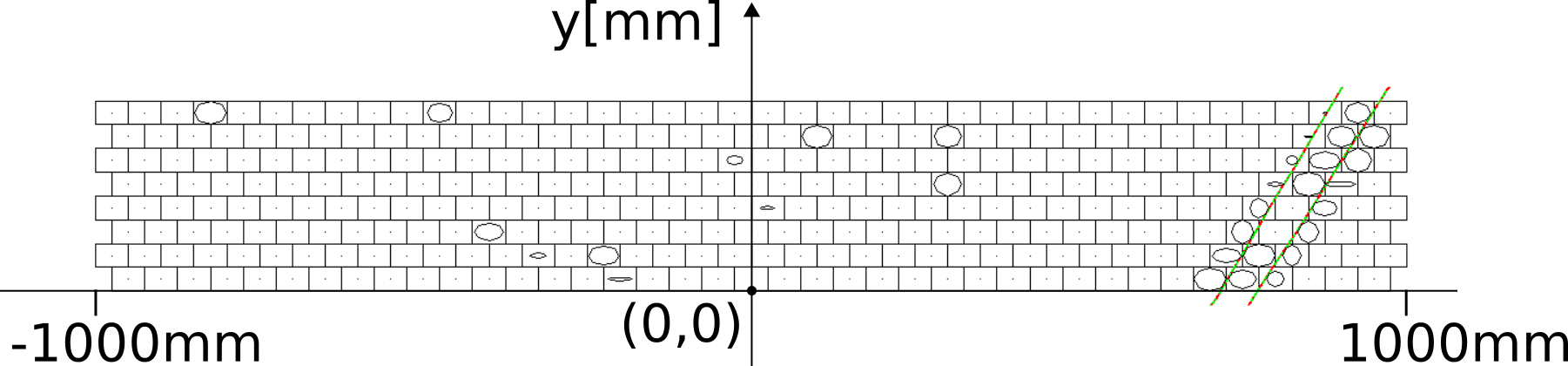}
        \caption{A dual track event with $50\%$ noise. Our method is robust against noise induced hits and tracks that are too close to each other.}
	\label{fig:dual_track_ex_robust}
\end{figure}
\end{landscape}

\begin{table}[h!]
\begin{center}
\caption{\normalsize Efficiency of the algorithm for different amounts of Monte Carlo generated tracks per event.}
\vspace{1ex}
\begin{tabular}{c c c c}
\hline\hline
Number of tracks & $1$ & $2$ & $3$ \\\hline
Efficiency [\%] &
 $ 99.8 \pm 4.5 $ & 
 $ 98.8 \pm 3.3 $ &
 $ 98.9 \pm 3.3 $ \\ 
Fake rate  [\%] &
 $  0.0 \pm 0.0 $ & 
 $  1.0 \pm 0.2 $ &
 $  0.9 \pm 0.2 $ \\
Fail rate  [\%] &
 $  0.2 \pm 0.1 $ & 
 $  0.1 \pm 0.1 $ & 
 $  0.1 \pm 0.1 $ \\
 \hline\hline
Standard deviation of relative error (slope) [\%] &
 $ \left(1.04 \pm 0.04 \right) \times 10^{-5} $&
 $ \left(1.16 \pm 0.03 \right) \times 10^{-5} $&
 $ \left(0.76 \pm 0.05 \right) \times 10^{-5} $ \\
Standard deviation of relative error (intercept) [\%] &
 $ \left(1.08 \pm 0.04 \right) \times 10^{-5} $&
 $ \left(1.12 \pm 0.03 \right) \times 10^{-5} $&
 $ \left(0.87 \pm 0.07 \right) \times 10^{-5} $ \\
Standard deviation of the residuals [$\mathrm{\upmu m}$] &
 $ \left(2.55 \pm 0.03 \right) \times 10^{-4} $&
 $ \left(2.65 \pm 0.02 \right) \times 10^{-4} $&
 $ \left(2.58 \pm 0.02 \right) \times 10^{-4} $ \\
  \hline\hline
\end{tabular}
\label{tab:multi_track_results}
\end{center}
\end{table}

\section{Conclusions}
\label{sec:conclusions}

We propose a pattern recognition method to detect the tangent lines of sets of ellipses using the Legendre transform.
This method reconstructs tracks of particles passing through a hypothetical detector. To evaluate our method, the reconstructed tracks are compared against the	 initial, Monte Carlo generated, tracks.
Based on the initial tracks, the ellipses with which the algorithm operates are created and then to simulate the measurement errors and the noise of the electronics, the smearing of the ellipses takes place and additional ellipses are created respectively.
The results show that our method is robust on detecting tangent lines of both ellipses with untampered characteristics and ellipses whose dimensions have been smeared, as well as with additional ellipses that simulate the noise of the electronics.


\begin{thebibliography}{00}

\bibitem{paper_legendre_sense} \href{http://scitation.aip.org/content/aapt/journal/ajp/77/7/10.1119/1.3119512} {R. K. P. Zia, \textit{et al.}}, Making Sense of the Legendre transform, American Journal of Physics, 77, 614-622, arXiv:0806.1147v2

\bibitem{paper_legendre_graph_deriv} \href{http://www.physics.drexel.edu/~skennerly/maths/Legendre.pdf} {Sam Kennerly}, A graphical derivation of the Legendre transform, 12 April 2011

\bibitem{paper_legendre_thermodynamics_calc} \href{http://scitation.aip.org/content/aapt/journal/ajp/72/6/10.1119/1.1648327} {Joel W. Cannon}, Connecting thermodynamics to students' calculus, American Journal of Physics, 72, 753-757

\bibitem{paper_legendre_thermodynamics_venn} \href{http://scitation.aip.org/content/aapt/journal/ajp/79/9/10.1119/1.3599177} 	{W. C. Kerr, J. C. Macosko}, Thermodynamic Venn diagrams: Sorting out forces fluxes, and Legendre transforms, American Association of Physics Teachers, 22 May 2011

\bibitem{paper_legendre_fenchel_computer} \href{http://www.doc.ic.ac.uk/~ahanda/lfreport.pdf} {A. Handa, \textit{et al.}}, Applications of Legendre-Fenchel transformation to
computer vision problems, \href{http://www.doc.ic.ac.uk/~ahanda/lfreport.pdf} {http://www.doc.ic.ac.uk/~ahanda/lfreport.pdf}

\bibitem{paper_legendre_circles} \href{http://www.sciencedirect.com/science/article/pii/S0168900208005780} {T. Alexopoulos, \textit{et al.}}, Implementation of the Legendre Transform for track reconstruction in drift tube chambers, Nuclear Instruments and Methods in Physics Research Section A: Accelerators, Spectrometers, Detectors and Associated Equipment, 592, 456-462

\bibitem{paper_legendre_circles_det} \href{http://www.sciencedirect.com/science/article/pii/S0168900214000977} {T. Alexopoulos, \textit{et al.}}, Identification of circles from datapoints using the Legendre transform, Nuclear Instruments and Methods in Physics Research Section A: Accelerators, Spectrometers, Detectors and Associated Equipment, 745, 16–23

\bibitem{atlas_cern} \href{http://iopscience.iop.org/article/10.1088/1748-0221/3/08/S08003/meta} {ATLAS collaboration} The ATLAS Experiment at the CERN Large Hadron Collider, Journal of Instrumentation, 3, S08003

\bibitem{lhc_cern} \href{http://iopscience.iop.org/article/10.1088/1748-0221/3/08/S08001/meta;jsessionid=F6C0E21090B696B7C19DB68F959DE8EA.c2.iopscience.cld.iop.org} {L. Evans, P. Bryant}, LHC Machine, Journal of Instrumentation, 3, S08001

\bibitem{paper_gaussian_sums} \href{http://arxiv.org/abs/1403.4413} {T. Alexopoulos, \textit{et al.}}, Identification of cirles from datapoints using Gaussian sums, arXiv:1403.4413v3



\end{thebibliography}
\end{document}